\documentclass[preprint]{aastex6}
\usepackage{amsmath}
\usepackage{selectp}
\usepackage{mathtools,amssymb}
\usepackage{multirow}
\usepackage[utf8]{inputenc}
\usepackage{nicefrac}
\usepackage{xfrac}
\usepackage{graphicx}
\usepackage{comment}

\usepackage[T1]{fontenc}

\newcommand\be{\begin{equation}}
\newcommand\ee{\end{equation}}

\newcommand{\ik}{{\it Kepler~}}

\def\kms{\ifmmode{\rm km\thinspace s^{-1}}\else km\thinspace s$^{-1}$\fi}

\begin{document}
%\tracingall
% ^this^ is what breaks it 

\pagenumbering{arabic}

%\slugcomment{Submitted to Astrophysical Journal}
\shorttitle{Visual Analysis of \ik TTVs}
\shortauthors{Kane et al.}

\title{Visual Analysis and Demographics of Kepler Transit Timing Variations}

\author{Mackenzie Kane\altaffilmark{1}, Darin Ragozzine\altaffilmark{1,2}, Xzavier Flowers\altaffilmark{1}, Tomer Holczer\altaffilmark{3}, Tsevi Mazeh\altaffilmark{3}, Howard M. Relles\altaffilmark{4}}

\email{darin\_ragozzine@byu.edu}

\altaffiltext{1}{Florida Institute of Technology, Department of Physics and Space Sciences, 150 West University Blvd., Melbourne, FL 32940, USA}
\altaffiltext{2}{Brigham Young University, Department of Physics and Astronomy, N283 ESC, Provo, UT 84602, USA}
\altaffiltext{3}{School of Physics and Astronomy, Raymond and Beverly Sackler Faculty of Exact Sciences, Tel Aviv University, Tel Aviv 69978, Israel}
\altaffiltext{4}{Harvard-Smithsonian Center for Astrophysics, 60 Garden Street, Cambridge, MA 02138, USA}

\setcounter{footnote}{0}

\begin{abstract} 
We visually analyzed the transit timing variation (TTV) data of 5930 Kepler Objects of Interest (KOIs) homogeneously. Using data from \citet{2014ApJ...784...45R} and \citet{2016ApJS..225....9H}, we investigated TTVs for $\sim$all KOIs in Kepler's Data Release 24 catalog. Using TTV plots, periodograms, and folded quadratic+sinusoid fits, we visually rated each KOI's TTV data in five categories. Our ratings emphasize the hundreds of planets with TTVs that are weaker than the $\sim$200 that have been studied in detail. Our findings are consistent with statistical methods for identifying strong TTVs \citep{2016ApJS..225....9H}, though we found some additional systems worth investigation. Between about 3-50 days and 1.3-6 Earth radii, the frequency of strong TTVs increases with period and radius. As expected, strong TTVs are very common when period ratios are near a resonance, but there is not a one-to-one correspondence. The observed planet-by-planet frequency of strong TTVs is only somewhat lower in systems with 1-2 known planets (7 $\pm$ 1 \%) than in systems with 3+ known planets (11 $\pm$ 2 \%). We attribute TTVs to known planets in multi-transiting systems, but find $\sim$30 cases where the perturbing planet is unknown. Our conclusions are valuable as an ensemble for learning about planetary system architectures and individually as stepping stones towards more detailed mass-radius constraints. We also discuss Data Release 25 TTVs, $\sim$100 KOIs with Transit Duration and/or Depth Variations, and that the Transiting Exoplanet Survey Satellite (TESS) will likely find only $\sim$10 planets with strong TTVs. 

\end{abstract}

\keywords{planetary systems; planets and satellites: dynamical evolution and stability}

\maketitle

\section{Introduction}
\setcounter{footnote}{0}

By recording the dimming and brightening of a star as a prospective transit occurs, NASA's Kepler Space Telescope determines the times when planet candidates move in front of their parent star. When the \ik pipeline identifies a possible planet, it is given a ``Kepler Object of Interest'' (KOI) designation. Many KOIs are false positives (e.g., eclipsing binaries (EBs) or instrumental effects), but thousands are planetary candidates or confirmed planets. 

Transiting planets (and EBs) can be gravitationally influenced by a perturbing object (or a non-spherical primary) which causes the transiting planet to appear late or early. These non-Keplerian deviations from a perfectly periodic set of transits are called “transit timing variations" (TTVs), as predicted by \citet{2005MNRAS.359..567A}, \citet{2005Sci...307.1288H}, and others. The dominant cause of TTVs in Kepler exoplanets is the gravitational perturbation of another planet, such that TTVs can be used to measure planetary masses. If the perturbing planet is also transiting (see Section \ref{multis}), then the mass and radius estimates can be combined to determine the density with crucial implications for planetary composition, formation, evolution, and habitability. 

The final best analysis to infer the physical and orbital properties of transiting exoplanets is the use of a photodynamical model \citep{RH10}. This method is the only recourse in the case for planets whose individual transits are too weak to measure transit times, though the new ``Spectral Approach'' is a strong step in this direction \citep{2018ApJS..234....9O}. However, when TTVs can be measured, they can be analyzed independently from the light curve in order to constrain the mass of the perturbing planet. Therefore, TTV analyses are a valuable first step towards measuring masses, even if they will eventually be superseded by photodynamical models. 

TTVs can also be used to detect non-transiting planets by inferring an unseen perturber. Non-transiting planets are common in \ik systems and their frequency is related to the inclination distribution of planetary systems \citep{2016ApJ...821...47B}. Therefore, the demographics of TTVs also provides crucial insights into the architectures of planetary systems \citep[e.g.,][]{2014ApJ...789..165X,2018arXiv180209526Z}. 

Several efforts have been made to identify and characterize \ik TTVs. Two major TTV catalogs were produced by \citep[][, hereafter H+16]{2016ApJS..225....9H} and \citep{2015arXiv150400707R}. H+16 identified 260 KOIs with ``significant'' TTVs chosen using quantitative statistical metrics with some minor visual vetting. Given the importance and variety of TTVs, we decided to augment the quantitative identification of important TTV systems with a visual inspection of the TTV data for 5935 KOIs. The visual inspection enabled a check on the results of H+16 to make sure no system with interesting TTVs was missed because of any inadequacy in the statistical metrics.

In Section \ref{methods}, we describe our inspection and visual rating system for these TTVs. The goals of this visual investigation were to confirm the statistical method for finding significant TTVs, identify interesting TTV signals missed by H+16, understand the distribution of TTV significance, and to provide a basis for additional analyses. These goals were achieved with results described in Section \ref{results}. We combine our data with other properties of these planets to examine the demographics of TTVs as a function of period, radius, and other properties with application to the Transiting Exoplanet Survey Satellite (TESS). Since TTVs are caused by interacting planets, we also investigate what we learn about multi-transiting systems from TTVs in Section \ref{multis}. Our conclusions are summarized in Section \ref{conclusion}. In two appendices, we address topics tangential to the main analysis: TTVs identified in \ik Data Release 25 and systems with interesting Transit Duration and Depth Variations. 

\section{Methods} \label{methods}

\subsection{TTV Data Provenance}

Due to the revolutionary breadth, cadence, duty cycle, homogeneity, and precision of \ik data, practically all statistically significant TTV signals are found among KOIs. Therefore, we do not consider any other sources of TTV data, e.g., from ground-based or other sources. 

We combined transit timing data from two sources: H16 and \citep{2014ApJ...784...45R} as updated by \citet{2015arXiv150400707R}\footnote{Available at \url{https://exoplanetarchive.ipac.caltech.edu/docs/Kepler\_KOI\_docs.html}}. Both sets of transit times are based on all seventeen quarters of \ik PDC-MAP Long Cadence\footnote{TTVs derived from Kepler Short Cadence data are more precise, but at present there is no published catalog of Short Cadence TTVs.} data retrieved from MAST in 2013-2014. H+16 \citep[an update of][]{2013ApJS..208...16M} focused on TTVs for 2599 planet candidate KOIs that passed some basic tests described therein, but provided TTVs for 3164 KOIs. As a product of the \ik mission, the data from \citet{2015arXiv150400707R} was produced more automatically, homogeneously, and for 4914 KOIs (including large numbers of known false positives). Most of these overlap and we inspected a total of 5930 KOIs. Note that our analysis covers about twice as many KOIs as H+16. New KOIs and TTVs from \ik's Data Release 25 \citep{2017arXiv171006758T} are briefly examined in Appendix \ref{DR25}. 

\citet{2015arXiv150400707R} included KOIs that we inspected, but which are not present in the official KOI list maintained at the NASA Exoplanet Archive\footnote{\url{https://exoplanetarchive.ipac.caltech.edu/}}. These KOIs were attributed to duplicate data, false-alarms, and not actual transits as described in Table \ref{table:koi_exclude}. Some KOIs were identified based on only 1-2 transits (e.g., KOI-1274.01) where TTVs are not meaningful, so no plots or ratings were created. 

\begin{center}
    \begin{longtable}[c]{c c}
    \caption{KOIs Not Included in NASA Exoplanet Archive\label{table:koi_exclude}}
\\    \hline
    KOI Number & Reason For Exclusion \\ 
    \hline
    \endfirsthead
    
    \hline
    \multicolumn{2}{c}{Continuation of Table \ref{table:koi_exclude}}\\
    \hline
    KOI Number & Reason For Exclusion \\ 
    \hline
    \endhead
    
    \hline
    100.02 & False-Alarm \\
    310.01 & Duplicate of KOI-114.01 \\
    342.01 & Duplicate of KOI-46.01 \\
    433.03 & False-Alarm \\
    1033.01 & Duplicate of KOI-51.01 \\
    1177.02 & False-Alarm \\
    1737.01 & Duplicate of KOI-442.01 \\
    1737.02 & Duplicate of KOI-442.02 \\
    1737.03 & Duplicate of KOI-442.03 \\
    3160.01 & Not a Transit \\
    3170.01 & False-Alarm \\
    3183.01 & Not a Transit \\
    3519.01 & Duplicate of KOI-3153.01 \\
    4915.01 & Duplicate of KOI-4980.01 \\
    4916.01 & Duplicate of KOI-4982.01 \\
    4917.01 & Duplicate of KOI 6112.01 \\
    4918.01 & Duplicate of KOI 6121.01 \\
    4919.01 & Duplicate of KOI-5115.01 \\
   \end{longtable}
\end{center}

\subsection{Creating Plots for Visual Inspection}

The first step of the visual inspection is to generate useful visualizations of the TTV data. Although it is not statistically valid, when TTVs were available from both sources, we decided to combine their individual data sets (using different markers for both). It is possible that this slightly biased our estimates for KOIs with data from both sources by presenting twice as many data points as were actually obtained. We do not consider this a serious issue since, 1) the data almost always agreed, so there was no need to consider each separately; and 2) doubling the data is not likely to change the score by much: strong TTVs are still visibly very different from mild TTVs. 

Most TTV signals are well described by a sum of sinusoidal signals due to various perturbations \citep[e.g.,][]{2008ApJ...688..636N,2012ApJ...761..122L,2016ApJ...818..177A}. In most cases, H+16 and others found that TTVs are well-explained by polynomial or sinusoidal fits, the former often due to sinusoidal signals with periods long compared to the length of the dataset (4 years). We found that the most common TTV signal could be very well represented by the sum of a single sinusoid and a quadratic baseline. (Note that the inclusion of a quadratic means that the sinusoid can be more accurately described in the presence of long-period trends.) 

For each TTV signal, we performed a sinusoid+quadratic fit that we call ``SinePoly''. SinePoly performs a least-squares fit\footnote{Recent work by \citet{2016ApJ...820...39J} suggests that fits assuming that TTV uncertainties follow Student t distributions with 2 degrees of freedom would be better.} to all the TTV data (combined if both sources are available) that is optimized using the IDL \texttt{powell} routine. The parameters are limited to reasonable ranges (e.g., amplitudes smaller than total range of the TTVs) and several initial guesses are employed. These guesses are based on independent quadratic and sinusoid fits, using a Discrete Fourier Transform to identify the peak of the fit after the best fit polynomial is removed, and an independent attempt to fit all parameters simultaneously. The best fit is then used as the final SinePoly estimate. Testing showed that SinePoly worked well on fitting TTVs for a variety of KOIs, despite its weaknesses (e.g., no testing for convergence or plausibility). It also provided a preliminary TTV fit that could be evaluated visually. We report herein the parameters from the SinePoly fit which can be compared to similar fits by H+16. 

We now describe the plots that were used for visual inspection. Our goal was to provide ratings describing various properties of the TTV data. Described in more detail below, we rated five aspects of the TTV data.

\begin{itemize}
\item Cleanliness rating: frequency of outliers
\item SinePoly Fit rating: how well the SinePoly fit performed
\item Periodogram Peak Periodicity rating: how clearly the TTV data are periodic after folding on a period determined from a periodogram
\item SinePoly Peak Periodicity rating: how clearly the TTV data are periodic after folding on the period determined from SinePoly
\item Overall Interest rating: an overall rating of the strength/interestingness of the TTV signal, taking the other scores into account
\end{itemize}

\begin{figure}						 
    \begin{center}
    \includegraphics[width=6in]{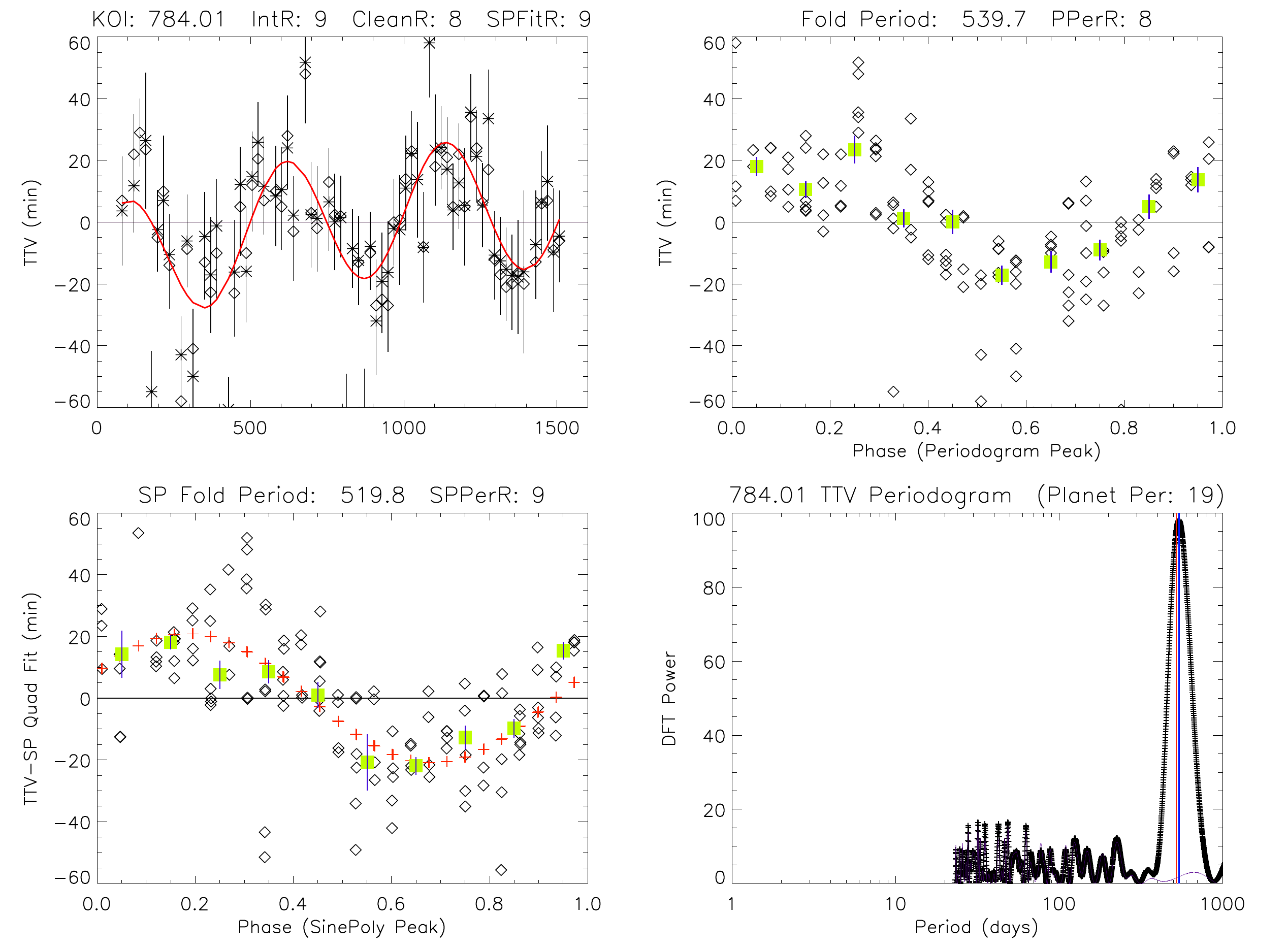}
	\caption{ Example of TTV data and plots that were visually inspected. Note that we have updated these plots to include our rating scores, so the plot shown here is different to the plots that were inspected (which has no effect on our conclusions). The TTV plot (upper left) shows the TTVs as a function of time, including uncertainties. To avoid the effect of outliers in controlling the y-axis range of the plot, we truncate the y-axis to the 95\% percentile or 4 times the median TTV uncertainty, whichever is greater. (This scale is preserved for all three plots that show TTVs.) Data from \citet{2015arXiv150400707R} are shown as stars and data from H+16 are shown as diamonds; if both are available, both are plotted, even though this may give more visual "weight" to signals that have both sources of data. (In other plots, all data are shown as diamonds.) The SinePoly sinusoid+quadratic fit is shown with red crosses (lower left) or a red line (upper left)  to help guide the eye. In the lower-right is the Periodogram plot, showing the power of the (combined, if available) TTV data determined using a Discrete Fourier Transform at 5000 points over a range from the orbital period of the planet to 1000 days. Black crosses and line show the periodogram of the original TTV data and a thin purple line shows the periodogram of the TTV data once the SinePoly fit has been removed. Except in cases of strong periodicity, a forest of low power peaks is generally seen. A blue vertical line shows the strongest periodicity and is the period used to phase-fold the TTVs for the Periodogram Peak plot (upper right). The red vertical line shows the SinePoly periodicity used in the SinePoly Peak plot (lower left). These often overlap. The Periodogram Peak plot and the SinePoly Peak plot each show phase-folded TTVs, with the quadratic component of the SinePoly fit subtracted from the latter. In both plots, the median value of the TTVs for 10 bins of width 0.1 in phase are shown by green squares. Squares are not plotted if there are fewer than three measurements in a bin. An error bar in the square shows 1.4826 times the Median Absolute Deviation of TTVs in that bin (chosen to represent a robust ``1-sigma'' error bar).}
	\label{fig:TTV_full}
    \end{center}
\end{figure}

In order to rate each KOI, the TTV data are presented in four panels, which we call the TTV plot, the Periodogram Peak plot, the SinePoly Peak plot, and the Periodogram plot. An example is shown in Figure \ref{fig:TTV_full}. All of our plots are available at \texttt{http://haumea.byu.edu/kanettv/\#.pdf} where \# is replaced with the KOI number, e.g., 784.01.pdf is at \url{http://haumea.byu.edu/kanettv/784.01.pdf}. (Note that these are not exactly the same as the plots used for visual inspection, as our ratings have been added in, but changes are not significant for the rating process.)

To aid in understanding all these plots, we explain them in detail here. The TTV plot shows the TTVs as a function of time, including uncertainties. Data from both sources are shown, when available, even though this may give more visual "weight" to signals that have both sources of data. The TTV plot includes a red line that shows the SinePoly fit. This plot is used to determine the Cleanliness Rating and the SinePoly Fit Rating. The inclusion of the SinePoly fit would help guide the eye, but the Overall Rating was not influenced by poor SinePoly fits. 

The Periodogram plot shows a periodogram computed using a Discrete Fourier Transform. Usually a forest of peaks, vertical lines are drawn at the location of the strongest periodicity and the period determined by SinePoly. No ratings are based exclusively on the Periodogram plot, but the presence of a dominant peak would influence the Overall Rating.

The Periodogram Peak plot and the SinePoly Peak plot are both similar, each showing a phase-folded version of the TTVs. The Periodogram Peak plot used a period for phase folding taken from the peak of the Periodogram, while the SinePoly Peak plot used the period determined by SinePoly. The Periodogram peak period was often subject to aliasing and the SinePoly period was almost always a more meaningful TTV period. Furthermore, in the SinePoly Peak plot, the quadratic fit from the SinePoly fit was removed, leaving only the periodic component. The SinePoly fit (now just a sinusoid) was also shown. 

In both of these plots, the phases were divided into 10 bins of width 0.1 and, if there were more than 2 datapoints in a bin, a square was placed at the median of the bin. An error bar in the square show a robust (based on the Median Absolute Deviation) ``1-sigma standard deviation.'' These ``median squares'' were important for assessing the validity of low Signal-to-Noise Ratio (SNR) variations. 

Due to an attempt to make all of the visual scores unbiased, any background information for the KOIs were not included in the data or researched while during the rating process. Only the KOI number and planet period were shown and no other information was used to influence the scores. In particular, the visual analysis was performed on all KOIs, regardless of their Kepler pipeline disposition as confirmed planets, planet candidates, or false positives. Many KOIs are eclipsing binaries (EBs), so the ``TTV'' curves are actually Eclipse Timing Variation curves (also known as O-C diagrams for ``Observed - Calculated''). We discuss the results for these systems separately. 

\subsection{Visual Inspection Rubrics}

Based on plots like Figure \ref{fig:TTV_full}, one of us (MK), visually inspected these TTV plots for all 5930 KOIs. Our goal was to validate the statistical methodology for identifying the strongest TTV signals and to potentially find interesting systems missed by other studies. Our goals were best served by using a homogeneous rating rubric which, by necessity, was devised before the data were deeply inspected. While our rubrics could be improved, the ratings developed therefrom were sufficient to answer our scientific questions. 

Each KOI was rated in each of the 5 categories on a scale from 0 to 9, although not all ratings effectively used the full dynamic range. The rubrics below describe how the ratings were assigned. These descriptions are qualitative and an improved understanding could be gained by comparing the TTV plots for KOIs with a variety of ratings. Inherent to the visual inspection is a subjective component that implies a natural uncertainty in any individual rating for a specific KOI. We discuss the distribution of these ratings and their correlation with other properties in Section \ref{results}. 

\subsubsection{Cleanliness Rating}
Transit times are determined by fitting transit light curve models to independent events. It is common for individual times to have poor fits for a variety of reasons (incorrect detrending, low SNR, wrong local minimum, etc.). Outliers are relatively common, so we used a Cleanliness rating to address how ``clean'' the data looked. This rating could be used to recommend KOIs where more detailed transit time fitting would produce better results. 

This rating was determined by noting what fraction of the TTV data error bars did not ``touch'' the SinePoly fit in the TTV Plot. Since we plotted ``1-sigma'' error bars, even the cleanest TTVs will not all touch the fit due to random noise. Still, assuming that the reported uncertainties are consistent, this helps to categorize all available KOI systems into those most likely to have clear signals versus those with significant outliers. If it appeared that 90\%+ followed a visible trend, then the system's TTV Plot received a score of a "9". If roughly 80\% followed the trend with a few outliers, the TTV Plot received a score of an "8". The score for the TTV Plot outliers decreases as the percent of data following the trend decreases. Note that the most egregious 5\% of outliers could be out of view, since the plot was trimmed to show only 95\% of the data. 

The score could also be reduced if the uncertainties were clearly underestimated. When the \citet{2015arXiv150400707R} and H+16 TTVs differed significantly, the cleanest curve was used. In plots with only a few TTV measurements, the SinePoly fit is able to match them very well, but this may not indicate that they are outlier-free. Such plots received a rating of "7". 

\subsubsection{SinePoly Fit Rating}
This rating gives a sense of how well and clearly the SinePoly fit matched the TTV data. If the SinePoly fit had very clear peaks and troughs, the TTV Plot received a "9". The SinePoly Fit rating would be an "8" if outliers interfered with a near perfect sinusoid. A rating of either a "7" or a "6" suggests that there may be a trend present, but to the eye it appears that there are multiple trends possible and it is undetermined if any are viable. The decreasing ranking of the cleanliness of the TTV Plot signal decreases with the unlikelihood of finding a noticeable signal. Only very poor fits received ratings less than 5. 

\subsubsection{Periodogram and Sinepoly Peak Periodicity Ratings}
The Periodogram Peak Periodicity rating and the SinePoly Peak Periodicity rating are determined using identical rubrics. These were based entirely off of the Periodogram Peak plot and the SinePoly Peak plot, respectively, with a focus on the aforementioned ``median squares'' that represent the phase-binned TTV measurements. 

A key criterion was to determine if there were the same number of median squares on either side of the TTV=0 horizontal line. If there are a substantial amount of median squares that appear to be sitting on the origin line with no visible sinusoidal fluctuation, the graph received a “5”. If there is either a small visible variation in the median squares that creates a low amplitude sinusoidal curve or a large variation of median squares that appears to be messy with a sinusoidal curve but a few median squares skewed from the curve, the graph received a “6”. If there is a clear variation of median squares that produces a clear sinusoidal curve but there are a few squares skewed a little to create a "bumpy" curve, the graph received a “7”. If there is a very clear variation of median squares and a very clear sinusoidal curve, the graph can be rated either an “8” or a “9” based on how clear the signal is. If there are too few median squares to assess the periodic nature of the TTVs, the KOI received a “0” rating.

\subsubsection{Overall Interest Rating}
Finally, an Overall rating was given that described the strength and interestingness of the TTV data based on all the plots. High ratings in the previous categories would increase the Overall rating. For example, a high Cleanliness rating and a high SinePoly Fit rating indicate that the SinePoly fit is very good, resulting in a higher Overall rating. If either (or both) Periodicity ratings found period with a very clear sinusoidal curve present, the Overall rating can be raised to an “8” based on how clear. The Periodogram plot would help identify the strength and uniqueness of these periodicities. This was particularly useful for TTV plots with many data points and short periods which were very difficult to interpret using just the TTV Plot. We note that a major goal of our analysis was to identify visually interesting TTV curves that did not fall under the usual categories of polynomials or sinusoids, and the overall interest rating allowed us to flag such systems as well, though they were rare.

\subsection{Additional Analyses of Individual Systems}

One of us (HR), analyzed some TTV data in greater detail, going well beyond the basic SinePoly fit. These analyses are not connected to our ratings, but provide additional information on many TTV signals, so we discuss them here. 

Hundreds of KOIs were inspected and the most promising investigated in further detail. Transit times were measured on the Kepler light curves obtained from the NASA Exoplanet Archive directly using the \texttt{EXOFAST} code \citep{2013PASP..125...83E}. TTVs were calculated in the usual way, i.e., by subtracting the best-fit linear ephemeris to the transit times. Lomb-Scargle periodograms were used to find the periodicities of each system using NASA's Exoplanet Archive Periodogram Service\footnote{\url{http://exoplanetarchive.ipac.caltech.edu/cgi-bin/Periodogram/nph-simpleupload}}. 

Best-fit linear+sinusoidal models were created using the Kaleidagraph\footnote{\url{http://www.synergy.com}} software package. When multiple models were adequate fits, preference was given to periods with stronger periodogram power. The best-fit parameters of these models are provided as discussed below. 

\subsection{Combined Exoplanet Data Table}

To further aid in the discovery of exoplanets, our SinePoly fit parameters, visual ratings, and more detailed TTV fit parameters are combined with statistical results from H+16, properties of the KOIs from NASA's Exoplanet Archive, and other sources. These have been compiled into a single large table, allowing us to analyze various correlations in Sections \ref{results} and \ref{multis}. 

The columns in the table are described in Table \ref{table:meta_data_table}. Due to the compilation of values from multiple different sources, some systems do not have values for some of the columns in the table. Therefore, the value of 99999 represents ``no value'' or ``Not Applicable''. This value also matches no other numbers found in the table and so can be treated unambiguously. 

\begin{center}

    \begin{longtable}[c]{r l l c}

    \caption{Columns of the Combined Exoplanet Data Table\label{table:meta_data_table}}
\\    \hline
    Column & Label & Description & Reference \\
    \hline
    \endfirsthead
    
    \hline
    \multicolumn{2}{c}{Continuation of Table \ref{table:meta_data_table}}\\
    \hline
    Column & Label & Description & Reference \\ 
    \hline
    \endhead
    
    \hline
1 & Name & KOI Number & \\
2 & SPPer & SinePoly best-fit period [day] & 1\\
3 & SPPhs & SinePoly best-fit phase [rad] & 1\\
4 & SPAmp & SinePoly best-fit amplitude [min] & 1\\
5 & SPLin & SinePoly best-fit linear coefficient [min/day] & 1\\
6 & SPQuad & SinePoly best-fit quadratic coefficient [min/day2] & 1\\
7 & SPChi2 & SinePoly best-fit chi-square & 1\\
8 & KIC & Kepler Input Catalog number (aka KID) & 2\\
9 & KepNm & Kepler planet name & 2\\
10 & EADisp & Exoplanet Archive Disposition & 2\\
11 & ClRat & Cleanliness visual rating (9 = fewest outliers) & 1\\
12 & SPFRat & SinePoly Fit visual rating (9 = outstanding fit) & 1\\
13 & PPRat & Periodogram Peak visual rating (9 = very strong periodicity) & 1\\
14 & SPPRat & SinePoly Peak visual rating (9 = very strong periodicity) & 1\\
15 & OvRat & Overall Interest visual rating (9 = very strong TTV signal) & 1\\
16 & Mult & Multiplicity = number of candidate/confirmed planets in system & 1\\
17 & KOIPer & KOI orbital period [day] & 2\\
18 & E\_KOIPer & KOI orbital period upper uncertainty [day] & 2\\
19 & e\_KOIPer & KOI orbital period lower uncertainty [day] & 2\\
20 & KOIT0 & KOI transit epoch (Barycentric Kepler Julian Date) [BKJD] & 2\\
21 & E\_KOIT0 & KOI transit epoch upper uncertainty [day] & 2\\
22 & e\_KOIT0 & KOI transit epoch lower uncertainty [day] & 2\\
23 & KOIDur & KOI transit duration [hr] & 2\\
24 & E\_KOIdur & KOI transit duration upper uncertainty [hr] & 2\\
25 & e\_KOIdur & KOI transit duration lower uncertainty [hr] & 2\\
26 & KOIror & KOI planet/star radius ratio & 2\\
27 & E\_KOIror & KOI planet/star radius ratio upper uncertainty & 2\\
28 & e\_KOIror & KOI planet/star radius ratio lower uncertainty & 2\\
29 & Rad & KOI planet radius [Rgeo] & 2\\
30 & E\_Rad & KOI planet radius upper uncertainty [Rgeo] & 2\\
31 & e\_Rad & KOI planet radius lower uncertainty [Rgeo] & 2\\
32 & KOISES & KOI maximum Single Event Statistic (similar to SNR of a single transit)  & 2\\
33 & KOIMES & KOI maximum Multiple Event Statistic (similar to SNR of all transits combined) & 2\\
34 & STeff & Stellar effective Temperature [K] & 2\\
35 & E\_STeff & Stellar effective Temperature upper uncertainty [K] & 2\\
36 & e\_STeff & Stellar effective Temperature lower uncertainty [K] & 2\\
37 & SMass & Stellar mass [solMass] & 2\\
38 & E\_SMass & Stellar mass upper uncertainty [solMass] & 2\\
39 & e\_SMass & Stellar mass lower uncertainty [solMass] & 2\\
40 & KepMag & Target magnitude in Kepler bandpass [mag] & 2\\
41 & PlMass & Planet mass [Mjup] & 3\\
42 & E\_PlMass & Planet mass upper uncertainty [Mjup] & 3\\
43 & e\_PlMass & Planet mass lower uncertainty [Mjup] & 3\\
44 & TTVErr & Median uncertainty on TTV measurements [min] & 4\\
45 & TTVSct & Robust scatter of TTV measurements [min] & 4\\
46 & p-Rat & log$_{10}$(p-value) for the Ratio of TTVSct/TTVErr & 4\\
47 & H16Per & Period of the highest TTV periodogram peak [day] & 4\\
48 & H16Pow & Power of the highest TTV periodogram peak [day2] & 4\\
49 & p-PS & log$_{10}$(p-value) for the highest TTV periodogram peak & 4\\
50 & H16Alm & Alarm score for TTV series & 4\\
51 & p-Alm & log$_{10}$(p-value) for the TTV Alarm score & 4\\
52 & H16pd & Polynomial degree chosen for TTV fit & 4\\
53 & p-poly & log$_{10}$(p-value) for the TTV Polynomial fit & 4\\
54 & HRLSP & Peak of the Lomb-Scargle Periodogram for independent TTVs [day] & 5\\
55 & HRPer & Period from the independent TTV sinusoidal curve-fit [day] & 5\\
56 & e\_HRPer & Uncertainty in period from the independent TTV sinusoidal curve-fit [day] & 5\\
57 & HRAmp & Amplitude from the independent TTV sinusoidal curve-fit [min] & 5\\
58 & e\_HRAmp & Uncertainty in amplitude from the independent TTV sinusoidal curve [min] & 5\\
59 & KOIprt & KOI attributed to the cause of the TTV perturbations & 1\\
60 & f\_KOIprt & Flag describing result in determining perturbing KOI & 1\\
61 & TDPV & Flag describing Transit Duration/Depth Variations & 1\\
62 & numH16 & Number of Transit Times from \citet{2016ApJS..225....9H} & 1\\
63 & numRT & Number of Transit Times from \citet{2015arXiv150400707R} & 1\\

   \end{longtable}
   
\tablecomments{The Combined Exoplanet Data Table is available as a Machine Readable Table.}
\tablerefs{
(1) This Work \quad
(2) NASA Exoplanet Archive (Cumulative KOI Table) downloaded June 6, 2016 \quad
(3) NASA Exoplanet Archive (Confirmed Planets Table) downloaded June 6, 2016 \quad
(4) H+16 \citep{2016ApJS..225....9H} \quad
(5) This Work \quad
}
\end{center}

\section{Distribution of Visual Ratings and Implications}  \label{results}

With our 5 ratings of 5930 KOIs, combined with auxiliary information from other sources, we can explore the connection between our ratings and the statistical metrics of H+16 and the demographics of TTV signals. We apply these results to topics related to multi-transiting systems in Section \ref{multis}.

\subsection{Distributions of Visual Ratings}

The distribution of our ratings for the 5930 KOIs inspected is shown in Table \ref{table:distribution_table}. As discussed above, since the rating rubrics were created before large numbers of KOIs were scored, some weaknesses in the results remain. In particular, the ratings did not utilize the full dynamic range from 0-9 and instead are mostly concentrated between 5-9. However, inspection of plots with a variety of ratings confirms that the general trends are robust and meet the desired goals of our analysis. 

\begin{center}
    \begin{longtable}[c]{c c|r r r r r}
    \caption{Distribution of Visual Ratings \label{table:distribution_table}}
\\    \hline
     Which KOIs? & Rating & Cleanliness & SinePoly Fit & Periodogram Peak & SinePoly Fit Peak & Overall \\
    \hline
    \endfirsthead
    
    \hline
    \multicolumn{2}{c}{Continuation of Table \ref{table:distribution_table}}\\
    \hline
    Which KOIs? & Rating & Cleanliness & SinePoly Fit & Periodogram Peak & SinePoly Fit Peak & Overall \\ 
    \hline
    \endhead
    
    \hline
    Planets & 0 & 4 & 6 & 408 & 456 & 8 \\
    Planets & 1 & 0 & 3 & 0 & 0 & 3 \\
    Planets & 2 & 1 & 3 & 1 & 1 & 9 \\
    Planets & 3 & 12 & 16 & 3 & 2 & 29 \\
    Planets & 4 & 141 & 92 & 6 & 2 & 45 \\
    Planets & 5 & 2096 & 1128 & 370 & 473 & 942 \\
    Planets & 6 & 872 & 1733 & 1941 & 1678 & 1745 \\
    Planets & 7 & 490 & 512 & 779 & 922 & 657 \\
    Planets & 8 & 139 & 186 & 156 & 155 & 190 \\
    Planets & 9 & 31 & 107 & 122 & 97 & 108 \\
    \hline
    
    All & 0 & 11 & 18 & 998 & 1079 & 23 \\
    All & 1 & 2 & 8 & 2 & 1 & 13 \\
    All & 2 & 6 & 17 & 4 & 4 & 28 \\
    All & 3 & 46 & 43 & 9 & 4 & 63 \\
    All & 4 & 405 & 253 & 16 & 8 & 281 \\
    All & 5 & 2972 & 1917 & 783 & 956 & 1654 \\
    All & 6 & 1298 & 2411 & 2585 & 2234 & 2419 \\
    All & 7 & 953 & 837 & 1095 & 1253 & 1011 \\
    All & 8 & 196 & 290 & 263 & 248 & 300 \\
    All & 9 & 42 & 136 & 175 & 143 & 138 \\
   \end{longtable}
\end{center}
\tablecomments{The distribution of our five visual ratings for the 3786 planets or planet candidates (based on the disposition in the NASA Exoplanet Cumulative Table on June 6, 2016) and all 5930 KOIs.}

Table \ref{table:distribution_table} shows that most ratings have a similar distribution, with 2-3\% having a score of "9" (exceptional), 5\% a score of "8" (strong), 15\% a score of "7" (weak/possible), and the remaining $\sim$75\% indicating very weak or indiscernible signals. Similar distributions are expected since each rating correlates strongly with the SNR of a TTV signal. The Periodicity ratings both show $\sim$1000 KOIs with a rating of "0" which resulted from long-period planets (or aliases) without enough TTVs to ascertain periodicity (too few ``median squares''). 

The Overall rating was the most important category, and will be considered in most detail below. A \emph{post facto} assessment of the Overall ratings suggests that "9" corresponds to the Strongest TTV signals; an "8" corresponds to a Strong signal; a "7" corresponds to a weak, minimal, and/or noisy signal; and "6" or below indicate no TTV signal of interest. For example, the H+16 list of interesting signals contained practically all "9"s and most "8"s. Many of the planets with an Overall rating of 9 have been analyzed directly in various TTV studies. 

Of particular interest is the 657 planets with an Overall rating of "7". More detailed and careful analyses of these systems should reveal large numbers of weak constraints on masses, densities, and/or the presence of non-transiting planets. Analyzing these as an ensemble could provide significant scientific insights, even though each individual measurement is weak. 

\subsection{Eclipsing Binaries}

Though EBs received the same treatment as the rest of the KOIs, they are not the focus of our study. We therefore removed these from Table \ref{table:meta_data_table} and discuss them here. Only known EBs are removed; it is likely that many additional unknown EBs remain and, indeed, our TTV plots can sometimes be used to help identify such false positives. 

We gather some of the properties of these EBs from Villanova's Kepler Eclipsing Binary Catalog\footnote{\url{http://keplerebs.villanova.edu/}} on June 6, 2016. These are combined with our ratings in Table \ref{table:eb_table}. These are provided as a basic reference; there are many other sources for determining EBs with interesting timing variations \citep[e.g.,][]{2016AJ....151...68K}.

\begin{center}
    \begin{longtable}[c]{r r r r r c c c c c}

    \caption{Analyzed Eclipsing Binaries\label{table:eb_table}}
\\    \hline

    KOI Number & KIC & Period (d) & K Mag & $T_{eff}$ (K) & Cleanliness & SinePoly Fit & Per. Peak & SinePoly Peak & Overall \\

    \hline
    \endfirsthead
    
    \hline
    \multicolumn{2}{c}{Continuation of Table \ref{table:eb_table}}\\
    \hline
    KOI Number & KIC & Period (d) & K Mag & $T_{eff}$ (K) & Cleanliness & SinePoly Fit & Per. Peak & SinePoly Peak & Overall \\ 
    \hline
    \endhead
    
    \hline

 225.01 &   5801571 &    1.7 & 14.78 & 6037 & 7 & 8 & 6 & 6 & 8 \\
1351.01 &   6964043 &    5.4 & 15.61 & 5374 & 6 & 6 & 6 & 7 & 6 \\
1452.01 &   7449844 &    1.2 & 13.63 & 6834 & 5 & 6 & 8 & 7 & 7 \\
1701.01 &   7222086 &    3.3 & 11.04 & 7065 & 6 & 6 & 7 & 6 & 6 \\
1771.01 &  11342573 &   91.1 & 15.96 & 5844 & 5 & 5 & 7 & 6 & 5 \\
3175.01 &   4909707 &    2.3 & 10.69 &   -1 & 7 & 8 & 8 & 8 & 8 \\
3244.01 &   6850665 &  214.7 & 12.39 & 4828 & 6 & 5 & 6 & 6 & 5 \\
3272.01 &   4948730 &   23.0 & 14.80 & 5624 & 8 & 8 & 8 & 7 & 8 \\
3290.01 &   4936990 &   10.3 & 15.29 & 6099 & 5 & 5 & 6 & 0 & 5 \\
3331.01 &   5876805 &   18.2 & 15.93 & 5559 & 5 & 6 & 7 & 6 & 6 \\
3467.01 &   7127885 &   33.9 & 15.19 & 5970 & 5 & 5 & 7 & 7 & 6 \\
3565.01 &   9592575 &    2.6 & 15.90 & 5436 & 7 & 8 & 7 & 8 & 8 \\
3606.01 &  10275074 &    4.4 & 14.18 & 6354 & 5 & 5 & 5 & 5 & 5 \\
3715.01 &   4937143 &    9.8 & 16.35 & 6705 & 8 & 9 & 9 & 9 & 9 \\
4294.01 &   7681230 &    1.0 & 14.84 & 6029 & 5 & 5 & 6 & 6 & 5 \\
4351.01 &   5436161 &    0.6 & 15.00 & 5001 & 5 & 5 & 6 & 6 & 5 \\
4925.01 &   1725193 &    5.9 & 14.50 & 5802 & 5 & 6 & 7 & 6 & 6 \\
4936.01 &   2305543 &    1.4 & 12.54 & 5623 & 5 & 6 & 6 & 7 & 6 \\
4953.01 &   2711123 &    0.7 & 12.53 & 4723 & 7 & 8 & 8 & 7 & 8 \\
4970.01 &   3245638 &    0.7 & 13.12 & 5883 & 5 & 6 & 6 & 6 & 6 \\
5015.01 &   3848919 &    1.0 & 13.90 & 5226 & 4 & 5 & 5 & 5 & 5 \\
5025.01 &   3953106 &   13.2 & 14.04 & 5398 & 4 & 6 & 6 & 6 & 6 \\
5061.01 &   4455763 &    0.8 & 15.58 & 6059 & 8 & 9 & 9 & 8 & 9 \\
5076.01 &   4732015 &    0.9 & 10.15 & 4185 & 7 & 8 & 9 & 9 & 9 \\
5090.01 &   4815612 &    3.9 & 15.18 & 6387 & 5 & 6 & 6 & 7 & 7 \\
5111.01 &   4996558 &    3.0 & 13.87 &   -1 & 5 & 5 & 6 & 6 & 5 \\
5112.01 &   5006817 &   94.8 & 10.87 & 4935 & 5 & 5 & 7 & 6 & 5 \\
5145.01 &   5263802 &    6.1 & 11.49 & 6642 & 9 & 9 & 9 & 9 & 9 \\
5152.01 &   5308777 &    0.9 & 13.20 & 4705 & 6 & 6 & 6 & 5 & 6 \\
5171.01 &   5467126 &    2.8 & 12.37 & 4683 & 5 & 6 & 6 & 6 & 6 \\
5233.01 &   6058896 &    1.1 & 14.78 & 5583 & 5 & 5 & 7 & 6 & 6 \\
5293.01 &   6525196 &    3.4 & 10.15 & 5966 & 5 & 5 & 6 & 0 & 5 \\
5353.01 &   7107567 &    0.8 & 14.23 & 6897 & 9 & 9 & 9 & 9 & 9 \\
5460.01 &   8016211 &    3.2 & 14.39 & 4933 & 6 & 7 & 7 & 7 & 7 \\
5564.01 &   8718273 &    7.0 & 10.56 & 4577 & 6 & 7 & 6 & 6 & 7 \\
5569.01 &   8747222 &    1.7 & 12.88 & 4777 & 5 & 6 & 7 & 6 & 6 \\
5683.01 &   9474485 &    1.0 & 14.88 & 4469 & 5 & 5 & 8 & 7 & 6 \\
5714.01 &   9786017 &    4.5 & 12.50 & 5753 & 5 & 5 & 7 & 6 & 5 \\
5733.01 &   9911112 &    2.3 & 14.99 & 8750 & 6 & 7 & 8 & 7 & 7 \\
5774.01 &  10191056 &    2.4 & 10.81 & 6588 & 7 & 7 & 7 & 6 & 7 \\
5797.01 &  10480952 &    4.1 & 12.22 &   -1 & 5 & 5 & 7 & 7 & 6 \\
5894.01 &  11401845 &    2.2 & 14.36 & 7590 & 7 & 7 & 7 & 6 & 7 \\
5906.01 &  11506938 &   22.6 & 11.61 & 6373 & 5 & 5 & 7 & 7 & 6 \\
5976.01 &  12645761 &    5.4 & 13.37 & 4844 & 6 & 7 & 8 & 8 & 7 \\

   \end{longtable}
\end{center}

The remaining discussions of TTV demographics \emph{use only systems identified as CANDIDATE or CONFIRMED in the Exoplanet Archive disposition as of June 6, 2016 (based mostly on DR24)}. We refer to these as ``planets'' throughout. Note that these dispositions sometimes exclude planets with very strong TTVs (see Appendix \ref{DR25}, but this is a small effect that we neglect.

\subsection{Data Comparison With H+16}
One of our major goals was to determine whether statistical methodologies worked well at identifying systems with strong TTVs. H+16 provide four statistical assessments of the strength of a TTV signal: the scatter p-value, the p-value of the F-test for the highest periodogram peak, the alarm score p-value, and the p-value of the F-test for the polynomial fit. See H+16 for more details. 

\begin{figure}						 
    \begin{center}
    \includegraphics[width=5in]{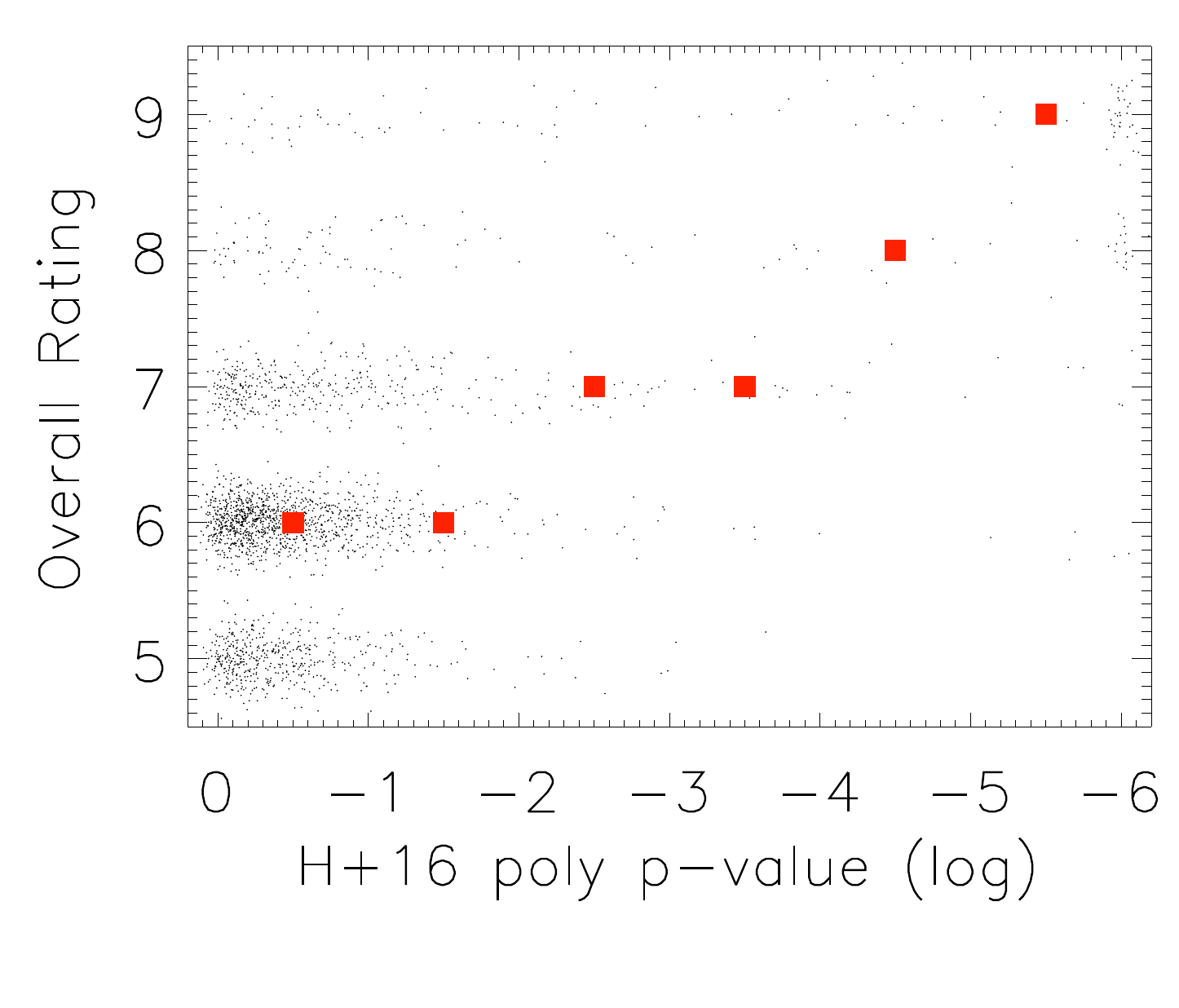}
	\caption{\label{fig:int_poly_scat} Data comparison between our visual Overall rating and the log$_{10}$ polynomial p-value estimate from H+16. The Overall rating accounts for more than polynomial signals, so this is not a direct comparison, but comparing to other H+16 p-values gives similar results. To aid the visualization, small random values are added to the Overall rating and the H+16 p-value and we do not display the small number of Overall ratings from 0-4. Also, p-values less than $10^{-6}$ are plotted at $10^{-6}$, explaining the pile-up on the right-hand side of the plot. Red squares show the median rating for planets in 1-dex-wide bins of the H+16 poly p-value. As expected, systems with no TTV signal, a weak TTV signal, a strong TTV signal, and the strongest TTV signals have typical Overall ratings of 6, 7, 8, and 9, respectively. Deviations from this trend have a reasonable explanation. For example, KOI-248.02/Kepler-49c has a strong sinusoidal signal that completes 4 full cycles over the course of the \ik data, so that there is no remaining polynomial signal. This is consistent with its position in the upper-left part of the diagram indicating a clear TTV signal, but no statistical preference for an overall polynomial trend. On the other hand, some KOIs with low p-values but low Overall ratings seem to have over-interpreted polynomial trends due to outliers.}
    \end{center}
\end{figure}

Our ratings and the H+16 p-values are not measuring the same quantities, but should show correlations. We find the expected correlations when considering: the F-test periodogram peak p-value vs. our Periodicity ratings, the alarm score p-value vs. our SinePoly Fit rating, the scatter p-value vs. our Overall rating, and the F-test polynomial p-value vs. our Overall rating. The last of these is shown in Figure \ref{fig:int_poly_scat} and is representative of the others. In all these comparisons, smaller p-values were correlated with higher ratings, as expected.

Figure \ref{fig:int_poly_scat} shows that practically all of the strongest signals (polynomial F-test p-values $\lesssim$10$^{-6}$) were given an Overall rating of 8 or 9. We confirmed through inspection that those low polynomial F-test p-values with high Overall ratings are merely indicative of a strong TTV signal that is not polynomial in nature. For each of the comparisons mentioned above, we investigated cases of disagreement between our ratings and the statistical p-values and found reasonable explanations for the disagreement. We conclude that both our ratings and the statistical metrics of H+16 are free of major issues. Furthermore, our visual ratings validate the statistical methods used by H+16 to select a subsample of the strongest TTV signals. 

Though our analyses agree in general, our work allowed us to identify several cases where H+16 missed KOIs with interesting TTVs. Three KOIs with an Overall rating of "9" were added to their list of interesting systems based on our analysis, including KOI-784.01 shown in Figure \ref{fig:TTV_full}. Twenty three planets with Overall ratings of "8" were not included in the H+16 list: 111.03, 117.01, 156.03, 232.04, 279.01, 351.04, 481.03, 899.03, 937.01, 1203.01, 1241.01, 1261.01, 1496.01, 1589.02, 1601.02, 1747.01, 1901.01, 2037.03, 2125.01, 2150.01, 3345.01, 3493.01, and 5605.01. The reasons they were not included vary: just below the ``interesting'' threshold, small amount of TTV data, outliers that confused the statistical metrics, or visually excluded. We reemphasize here that H+16 were not aiming for a rigorously complete sample when they chose the systems for a more detailed TTV analysis -- these ``missing'' KOIs are not a criticism of their work. 

In addition to these 23 planets, several other planets (and large numbers of false positives) with Overall ratings of 8 or 9 are not included in the H+16 list for various reasons. Our ratings and plots are one way to identify systems for future more detailed TTV analyses. 

\subsection{TTV Demographics}

\begin{figure}						 
    \begin{center}
    \includegraphics[width=5in]{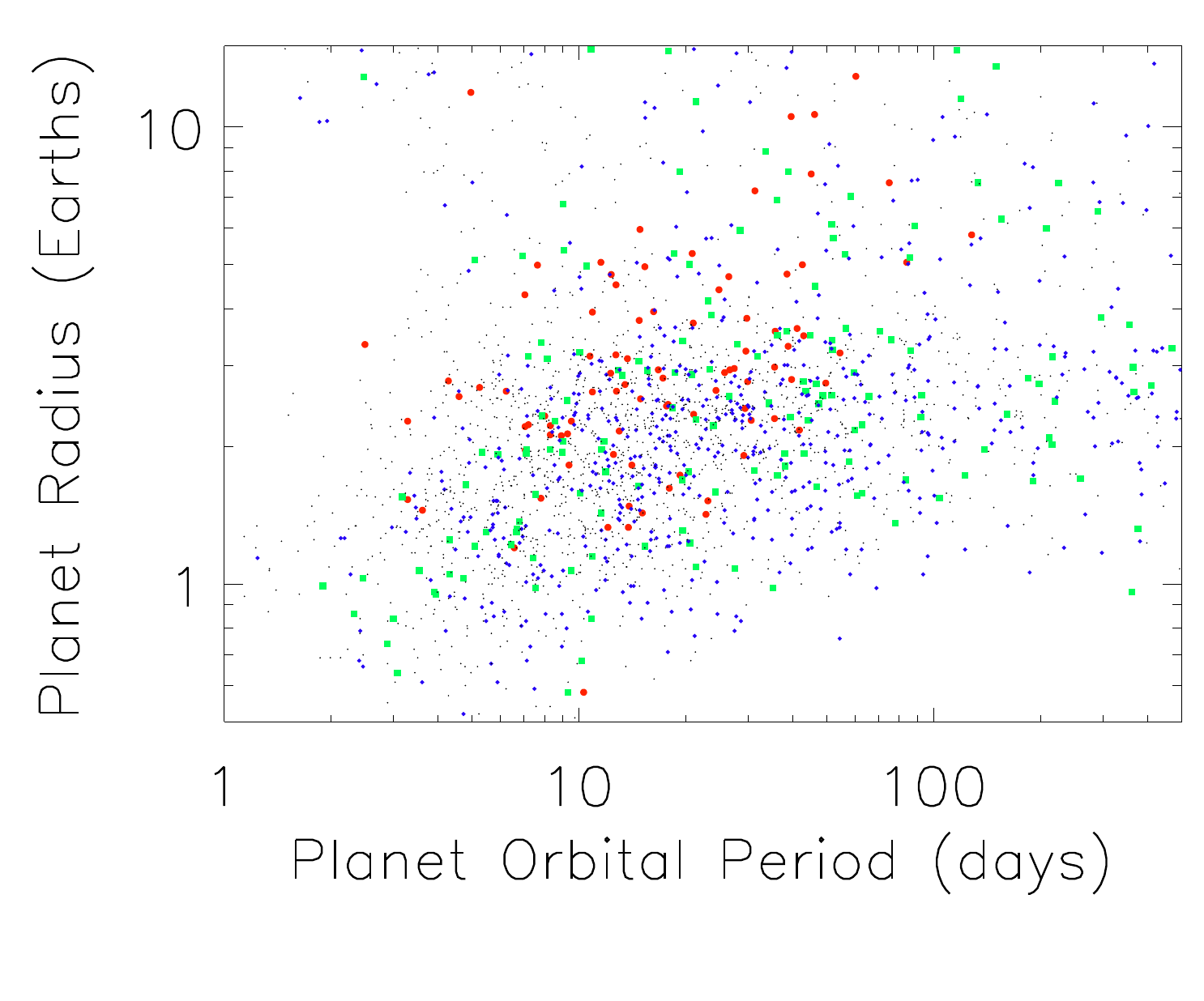}
	\caption{\label{fig:perradttv} The radius vs. period distribution for planets for different Overall ratings. As is customary, the period and radius axes are logarithmic and zoomed in on the region where most objects have been discovered. Planets with a rating of less than 7, 7, 8, and 9 are plotted with black dots, small blue diamonds, green squares, and red circles, respectively. For periods between 1-50 days and radii between 1.3-8 Earth radii, TTV strength increases with both period and radius. Of particular note is the dearth of strong TTVs with periods less than 3 days. The lack of the strongest TTVs at longer periods and smaller radii appears to be due to observational bias.}
    \end{center}
\end{figure}

Our classification of all KOIs allows us to consider the ``demographics'' of TTVs and how stronger TTVs correlate with other parameters. For example, Figure \ref{fig:perradttv} shows the period-radius distribution as a function of Overall rating, which illustrates trends discussed below. 

While the apparent strength of TTVs depends on many factors, we here report only on what is observed without an attempt to control for observational biases. The results in this section are only for KOIs which have Exoplanet Archive dispositions of CANDIDATES or CONFIRMED, as discussed above. Demographics related to multiple planets in the same system are discussed in Section \ref{multis}. Our results are generally consistent with \citep{2014ApJ...789..165X}. 

\subsubsection{TTVs and Orbital Periods}

Strong TTVs are rarely present around planets with orbital periods of $\lesssim$3 days. This is further emphasized when inspecting those systems with TTV Overall ratings of 8-9 and periods less than 3 days -- there is often evidence that these are actually eclipsing binaries (despite their candidate/confirmed disposition). Most of these systems have Kepler pipeline planet fits with planets several Jupiter radii and impact parameters greater than 1, a clear indication of a V-shaped transit characteristic of EBs. As there are many known EBs with periods less than 3 days that show strong ETVs \citep{2016AJ....151...68K}, this is not surprising.

Considering planets with radii <15 Earth radii (to help exclude EBs) shows that planets with an Overall rating of 7, 8, and 9 are all clearly different from the rest of the distribution because they lack planets less than $\sim$3 days. The lack of strong TTVs at periods $\lesssim$3 days is a very striking result. To check whether this is related to our visual analysis, we also inspected the four statistical metrics in H+16, where we also found that the strongest signals were very rare around short period planets and that exceptions had evidence for being false positives. It also seems to be present in \citet{2018ApJS..234....9O} who are sensitive to TTVs at shorter periods.

\citet{2012PNAS..109.7982S} already showed that Hot Jupiters (which are often in this period range) do not show strong TTVs\footnote{The only Hot Jupiter with an Overall rating of 9 is KOI-760.01 with a period of 5 days. It does not appear to be a blended eclipsing binary. The TTVs could potentially be caused by light-travel time variations due to a low-eccentricity massive brown dwarf with a period of 1050 days.}. Of great interest is that small short-period planets, the Hot Earths, also show very few TTVs. \citet{2016PNAS..11312023S} show that a significant fraction of these objects are not attributable to any known population and this is strongly supported by our result that TTVs are rare among such systems. More specifically, the lack of TTVs emphasizes that these Hot Earths have few nearby companions, transiting or not. In addition to the mechanisms discussed by \citet{2016PNAS..11312023S}, we suggest that tidal interactions may pull the innermost planet in fastest, preferentially separating it from any external companions \citep[e.g.,][]{2016AJ....152..105M}. Even a small (factor of $\lesssim$2) increase in the period ratio between hot planets and the next outer planet would be sufficient to minimize TTVs and to increase the relative geometric probability of detecting only one planet. 

There is an overall increase in the average TTV Overall rating as a function of period up to about 50 days. Strong TTVs are relatively rare among planets with periods $\gtrsim$50 days. This is understandable since such planets would only have $\sim$30 transits over the course of the \ik mission and identifying clear significant trends requires unusually large TTVs (though such cases exist). 

Like \citet{2013ApJS..208...16M}, we observe a weak correlation between the planet period and the TTV period for systems with detectable TTV signals. We also observe a weak correlation between the planet period and the TTV sinusoid amplitude.  

\subsubsection{TTVs and Planetary Radii}

TTVs with Overall ratings of 9 are very rare around planets with (Exoplanet Archive) radii $\lesssim$1.3 Earth radii. In addition, the average Overall rating increases with radius (up to $\sim$6 Earth radii). Such a trend is expected: smaller planets have weaker transits and thus higher TTV uncertainties, making strong TTVs harder to recognize. It is important to note that, as discussed in Section 1, many small planets are missing from our sample because they may not have individual transits strong enough to measure a reliable transit time, which is a prerequisite to making our TTV plots and ratings. 

We can assess the effect of observational bias by considering the distribution of Overall ratings with respect to similar measures: planet radius over star radius (closely related to transit depth), transit duration, maximum Single Event Statistic (closely related to SNR-per-transit), Multiple Event Statistic (closely related to total SNR), TTV median error (from H+16), and Kepler magnitude (related to the photometric uncertainty). In each case, the result is consistent with the strongest TTV signals being suppressed by observational uncertainty. Furthermore, there is a strong negative correlation between TTV sinusoid amplitude and Single and Multiple Event Statistic, which is almost erased when comparing the TTV sinusoid amplitude as a function of planet radius. Therefore, the lack of visibly strong TTVs for the smallest planets is consistent with an observational bias. 

The Overall rating correlates with radius up to about 6 Earth radii. In the jovian planet regime, there are systems with strong TTVs, but it is conceivable that observational bias inflates their importance and that the intrinsic frequency of TTVs is smaller for larger planets. That is, due to their high SNR (and potentially massive perturbers), the lower incidence of strong TTVs indicates that such systems may have different architectures. 

We have identified some empirical trends in the observability of TTVs as a function of period and radius which we consider as starting points for future more rigorous analyses.

\subsubsection{TTVs and Stellar Properties}
A first glance at the distribution of stellar effective temperatures (from the Exoplanet Archive) shows an unusual enhancement of strong TTVs near 3700-4000 K followed by a dearth between 4000-5000 K. However, inspection shows that the enhancement is due to multiple planets showing TTVs in a single system. Considering the Overall rating in stellar temperature bins, we find no correlation, consistent with the hypothesis that low-mass stars have more planets with TTVs only because they have higher multiplicities.

\subsection{Implications for TESS}

The Transiting Exoplanet Survey Satellite (TESS) is an upcoming space-based transit survey with a major goal of improving the mass-radius relation of small planets \citep{2014SPIE.9143E..20R}. Kepler's ability to identify many masses using TTVs \citep[e.g.,][]{2017AJ....154....5H} suggests that TTVs could play a major role in this fulfilling TESS's goals. 

However, the detectability of TTVs is strongly enhanced by acquiring long time-baselines. Kepler's four-year duration has revealed many TTV signals that would be missed by TESS, which has a one month duration for most of the sky and one year at most in its continuous viewing zones near the ecliptic poles. Still, the first Kepler TTVs around Kepler-9 were clearly identified with only seven months of data \citep{2010Sci...330...51H}. 

Based on our Kepler results, how many TESS systems might show TTVs?\footnote{\citet{2015ApJ...809...77S} simulate the number of planets that TESS is expected to detect based on Kepler results for planet frequency and estimated properties of TESS targets. While they nominally included multi-planet systems, it was not designed to accurately represent the architectures of these systems. As a result, they predicted that basically no multis would be detected by TESS, though comparison to actual Kepler systems indicates that this is due to a period ratio distribution biased towards much larger period ratios. \citet{2018arXiv180104949B} shows that using accurate architectures would result in the detection of many multiply-transiting systems, i.e., analogs of Kepler systems where multiple transiting planets are seen with periods of several days. The fact that \citet{2015ApJ...809...77S} did not expect to detect multi-planet systems should not be taken as an argument that TESS will not see systems with TTVs.} A proper answer would involve rigorous simulations of accurate planetary systems, including the correlations with near-resonant period ratios, an estimated mass-radius-eccentricity relation, careful signal-to-noise calculations, and other complications. An excellent first step towards this analysis was recently given by \citet{2018arXiv180104949B}. Here, we focus on a simpler problem of asking how TTV signals degrade when the observing duration goes from Kepler's four year long mission to TESS's durations, which was not considered explicitly by \citet{2018arXiv180104949B}. 

We do not expect any planetary TTVs to be detected over most of the sky where TESS only observes for one month. At best, TESS observes for a full year. We took the Kepler planets with an Overall rating of 9 (strongest TTVs) and generated our TTV plots\footnote{available at \url{http://haumea.byu.edu/kanettv/TESS\_TTV\_1year.pdf} and \url{http://haumea.byu.edu/kanettv/TESS\_TTV\_6mo.pdf}.} based on only the first year of data (after subtracting out the best-fit line to account for the unknown orbital period). As expected, the periodogram and SinePoly fits were poorer and more strongly affected by noise in these shorter datasets. Even so, these systems were then rescored using the same rubric in Section \ref{methods}. 

As expected, most systems became undetectable when only one year of data was available. However, $\sim$30\% of systems still showed strong TTVs (scores of 8-9) and $\sim$30\% have some evidence of TTVs (scores of 7). Interestingly, readily detectable TTV systems are still possible even with TESS's cadence; approximately 2\% of all Kepler planets have TTV signals with variations fast enough to show deviations from a linear ephemeris within one year. Even with only six months of TTV data, $\sim$10\% of these systems retained strong TTVs (scores of 8-9) and $\sim$30\% had indications of TTVs (score of 7). Given the similarity to early Kepler results \citep{2011ApJS..197....2F} that many planetary systems are known where the TTV period is around a year, this is not too unexpected. 

We make a very rough estimate of TESS's ability to detect TTVs. Using results from \citet{2015ApJ...809...77S} and the assumptions that 1) TTVs are most common around sub-jovian planets; 2) about half of TESS  planets will be detected in the regions with six or twelve month durations; and 3) $\sim$2\% of planets will show detectable TTVs, we can estimate that TESS will detect $\mathcal{O}$(10) planets with TTVs. This analysis is supported by the detailed analysis of \citet{2018arXiv180104949B} which predict that $\sim$5\% of TESS planets will have TTVs of similar strength to Kepler and our result that about $\sim$30\% of these will have TTVs that can be characterized with TESS's shorter observation duration (giving $\sim$2\%). Only a fraction of these will yield reliable and decent mass estimates. Hence, we expect that TTVs will not significantly contribute to TESS's goal of measuring masses for small planets, though this conclusion should be taken lightly without a much more rigorous analysis. 

\citet{2017arXiv170508891B} consider the possibility of TESS extended missions and find a trade-off between extending data coverage of stars observed in the main mission and observing new fields. Longer observations of known planets enable significantly improved TTV analyses, supporting plans to obtain continuing data \citep[see also][]{2013arXiv1309.1177F}.

\section{TTVs in Multi-Transiting Systems} \label{multis}
The results discussed so far have been independent of the number of known transiting planets (multiplicity), consistent with the findings of \citet{2014ApJ...789..165X} and \citet{2018arXiv180209526Z}. We now turn to the correlation between TTV strength and multiplicity that is expected as TTVs are caused by planet-planet interactions. Note that multiplicity was not considered when rating TTV systems (each system was rated individually with no consideration given to multiplicity). 

\subsection{TTVs and Observed Multiplicity}

Comparing the frequency of strong TTVs (overall scores of 8-9) and using basic Poisson uncertainties confirms the expected correlation with multiplicity, also seen by \citep{2014ApJ...789..165X}. The fraction of planets from singly and doubly-transiting systems with strong TTVs (8-9) is about 7 $\pm$ 1\% (on a planet-by-planet basis). The fraction of planets with strong TTVs from systems with 3 or more planets is about 11 $\pm$ 2\%. This is true for 3, 4, and 5-7 planet systems, independently, within Poisson errors. It is interesting that the increase in TTV frequency clearly occurs between systems with 2 and 3 transiting planets and not between 1 and 2 transiting planets, though the results of \citet{2014ApJ...789..165X} indicate that this may depend on how the planet sample is defined. 

While strong TTVs are more prevalent among planets from systems with 3-7 planets, the prevalence of TTVs in singly-transiting systems is really not much different than in multi-transiting systems (on a planet-by-planet basis). Since singly-transiting systems compose 61\% of the planets investigated, though TTVs are slightly rarer in such systems, they still contain 53\% of the 298 planets with strong TTVs. 

The preference for strong TTV scores in higher multiplicity systems is presumably due to these systems being more tightly packed (whether or not all planets in the system are seen). On the other hand, mild TTVs (score of 7) are more prevalent around singly transiting systems (19 $\pm$ 1 \%) compared to multi-transiting systems (13 $\pm$ 2 \%), though biases make it difficult to ascribe much meaning to this finding. 

Since strong TTVs are associated with resonances (as discussed below), it is worth considering whether TTV scores are distributed randomly among systems or whether the presence of one planet with strong TTVs enhances the probability of a second planet with strong TTVs. To test this scenario, we scrambled TTV scores among all systems that had the same multiplicity and compared the average TTV score in each system to the real unscrambled case. Using a two-sample K-S test shows that these distributions are clearly different with a p-value of $\sim$0.003. Inspection of the results showed that planets with high TTV scores were clearly grouped into pairs, as expected.

\subsection{TTVs and Period Ratios}

As mentioned above, the TTVs were rated entirely individually with no reference to other planets in the systems. Therefore, there is no explicit bias in looking at TTV scores as a function of period ratio. 

\begin{figure}						 
    \begin{center}
    \includegraphics[width=5in]{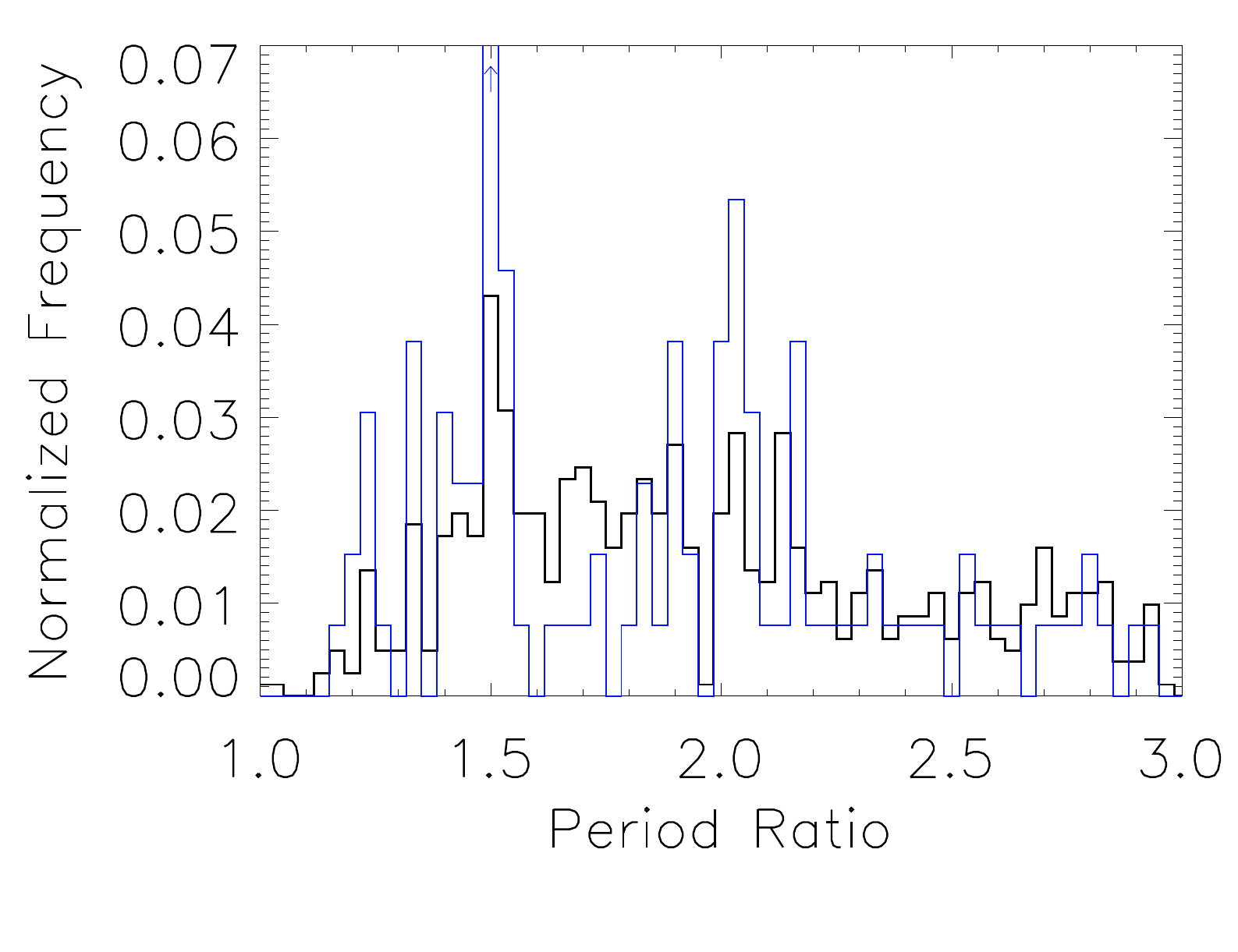}
	\caption{\label{fig:perratio} Comparison of the normalized period ratio distribution for all planets (thick black histogram) to the same distribution for planets where at least one of the planets in the pair has an Overall rating of 8 or 9 (thin blue histogram). Only neighboring period ratios are shown, though we cannot account for missing intermediate planets (see Section \ref{missing}). The peak for the frequency histogram for strong TTVs is located at a period ratio of 1.5 and is well off the plot as indicated by the upward arrow; the normalized frequency in this bin is 0.13. The histograms are normalized over all period ratios, but only period ratios from 1-3 are shown with a bin size of $\frac{1}{30}$. Changing the range or binsize does not affect our conclusions. The period ratio distribution with a ``continuum'' peaking at $\sim$2 and ``spikes'' just outside of resonances is well known \citep[e.g.][]{2011ApJS..197....8L, 2015MNRAS.448.1956S, 2016ApJ...821...47B}. As TTVs are more detectable by \ik for near-resonant period ratios, the strong-TTV period ratio distribution is even more peaked at these locations, as expected. However, there is not a one-to-one correspondence between near-resonance period ratios and strong TTVs. The plot has a similar character when considering the frequency of period ratios where both planets show strong TTVs.}
    \end{center}
\end{figure}

TTVs are known to be strongly enhanced near low-order commensurabilities due to the effects of (near-)mean motion resonances \citep[e.g.,][]{2012ApJ...761..122L}. We clearly see significant ($\sim$2x) enhancements in the frequency of strong TTVs in these regions on top of the mild (factor of $\sim$2) enhancement in the overall frequency of planets with near-resonance period ratios \citep{2011ApJS..197....8L}. Additional enhancement is seen for systems with period ratios $\lesssim$1.4. Many of these are near first-order resonances (4:3, 5:4, 6:5, and 7:6), but the general proximity of the interacting planets also is expected to increase TTVs. Figure \ref{fig:perratio} compares the normalized period ratio distribution for all systems to those with strong TTVs. 

While the expected correlations with resonances and small period ratios are seen, it is also seen that these are not complete predictors of strong TTV signals. There are many systems with strong TTVs that are not near resonances. Many, but not all, of these cases are presumably due to unknown intermediate planets (discussed in detail below). TTVs can help characterize a variety of planets, not just the near-resonant pile-ups. 

There are many near-resonant pairs without strong TTVs, but this is likely just an observational bias: we confirm that near-resonant pairs with strong TTVs have significantly higher planetary radii than pairs without strong TTVs. Since TTV amplitude correlates with mass (which correlates somewhat with radius) and TTV precision correlates with radius, such a finding is expected \citep[see also][]{2014ApJ...789..165X}.

There are multiple ways to quantify which periods are ``near-resonance'' (e.g., Steffen \& Hwang 2014), but \citet{2012ApJ...761..122L} show that $\Delta$ has dynamical meaning: 
\begin{equation}
\label{ttvdelta}
\Delta \equiv \frac{P_{\rm outer}}{P_{\rm inner}} \frac{q}{p} - 1 
\end{equation}
where the planets are in a $p:q$ resonance ($p > q$).\footnote{\citet{2012ApJ...761..122L} define $\Delta$ for first-order resonances where $q=p+1$ in our notation, but its application is valid in our generalized case.} Effectively the fractional offset from resonance, $\Delta$ also relates the planetary orbital period to the TTV period in the most common case of near-resonant planets. 

Near the 2:1 resonance, strong (overall score of 8-9) TTVs are prominent among period ratios of 2.02-2.08 (0.01 < $\Delta$ < 0.04), with strong TTVs relatively uniform over this range. Near the 3:2 resonance, strong TTVs are prominent among period ratios of 1.5-1.52 (-0.01 < $\Delta$ < 0.02), peaking sharply at $\Delta$ = 0.01. Including mild TTVs would extend these ranges somewhat to larger period ratios (but not smaller period ratios). These ranges match values of $\Delta$ for strong TTVs published in Table 8 of H+16. 
The difference between the $\Delta$ ranges with TTVs for these two major resonances could be related to other differences between these resonances (e.g., the 3:2 period ratio spike extends narrow and wide of the resonance, while there is a strong gap narrow of the 2:1 resonance). Other resonances also have enhancements, but their statistical significance is more difficult to establish.

\subsection{Missing Planets in Multi-Transiting Systems}
\label{missing}

The amplitude of a TTV signal scales with the mass of the perturbing planet and not the planet that is showing TTVs itself. This is why TTVs in multi-transiting systems are so valuable: if TTVs measure the mass of a perturbing planet that is also transiting, then its density can be obtained \citep{RH10}. TTVs on a planet in a singly-transiting system can provide some constraints, but even strong TTVs admit degeneracies in the perturbing planet's mass and period \citep[e.g.,][]{2010ApJ...718..543M,2011ApJ...743..200B}. With a few notable exceptions, most TTV analyses have focused on multi-transiting systems. 

However, there are many cases where TTVs are not caused by the other known transiting planets. For example, the KOI-884/Kepler-247 system has three known transiting planets at periods of 3, 9, and 20 days, but \citet{2014ApJ...790...31N} find that the TTVs on the 20-day planet are clearly caused by a non-transiting planet with a 60-day period.

H+16 identify which TTV signals can be associated with known planets (their Table 8) and we extend that work here. We attempt to identify which planet is causing TTVs for our strongest signals (Overall rating of 8-9) or if there are no good candidate perturbers. The perturbing KOI is identified by checking the period ratios, looking for TTVs with periods near the expected super-period \citep{2012ApJ...761..122L}, and checking for anti-correlated signals. In many cases, the nearest KOI is not within a period ratio of 2.5, indicating minimal dynamical interaction. There are a few cases where there is a nearby planet, but it does not seem to be the source of the TTVs. 

This search was augmented by students in the Fall 2015 ``Introduction to Planetary Science'' (SPS1020) class at the Florida Institute of Technology, under the direction of co-authors DR and MK. These students were taught about TTV signals and data fitting. They used the research-grade \emph{Systemic} software \url{http://www.stefanom.org/console-2/} to perform basic TTV fits, focusing on circular orbits with reasonable masses. \emph{Systemic} allows for one-dimensional and multi-dimensional minimization which was used to identify appropriate periods/phases and estimate masses, respectively. These analyses focused on finding global minima, but not on determining uncertainties. All multi-transiting systems where one or more planet in the system had a TTV rating of "7" or higher were investigated. Students checked each other's work and were also reviewed by co-authors DR, MK, and XF. Several TTV fits were consistent with published results. 

Using H+16 and augmented by these fits, we attempted to determine the perturbing planet for all objects with an Overall rating of 8-9. Like the visual ratings, this was done subjectively based on inspection of the systems and is likely to be correct in general but possibly incorrect for particular systems. Usually, the conclusion that TTVs were not due to the known planets is derived from very large period ratios between the planet with TTVs and the other planets in the system. In a few cases we choose to ascribe the TTVs to a non-transiting planet in preference to a poor or unlikely fit from a known planet. 

The results of our analysis are included in Table \ref{table:meta_data_table}. We include two columns: \texttt{KOIprt} is the KOI number of the planet identified as responsible for the TTVs (if any) and \texttt{f\_KOIprt} is given only one of the following values. Parentheses denote how many planets (not KOIs) have the corresponding \texttt{f\_KOIprt}.

\begin{itemize}
\item -1 $\rightarrow$ Overall rating of 7 or below (3197)
\item 0 $\rightarrow$ Overall rating of 8-9, but not in a multi-transiting system (122)
\item 1 $\rightarrow$ TTVs clearly caused by the planet listed in \texttt{KOIprt} (67)
\item 2 $\rightarrow$ TTVs probably caused by the planet listed in \texttt{KOIprt} (11)
\item 3 $\rightarrow$ TTVs could be caused by the planet listed in \texttt{KOIprt} (16)
\item 4 $\rightarrow$ TTV signal inconsistent with neighboring planets (8)
\item 5 $\rightarrow$ no planet within period ratio of 2.5 (27)
\item 6 $\rightarrow$ doesn't fit into any of the above categories (2)
\end{itemize}

Our results show that about 1/4 of TTVs in multis cannot be easily ascribed to the known planets (e.g., there are 37 planets with \texttt{f\_KOIprt} > 3 compared to 94 planets with \texttt{f\_KOIprt} $\le$ 3). Inspection of these systems shows a trend consistent with perturbers that are not known primarily because they are not transiting (as opposed to below the detection threshold). For example, as multiplicity increases, the number of inexplicable TTV signals drops significantly. Geometric probability requires that multi-planetary systems should regularly have non-transiting planets \citep{RH10,2016ApJ...821...47B}. Though planets with higher semi-major axes are more likely to not transit, it is not uncommon for planets between known transiting planets to be non-transiting \citep[e.g.,][]{2016AJ....152..160B}. Hence, our identification of many $\sim$30 strong TTV signals in multis that cannot be ascribed to the known planets is consistent with expectations for non-transiting planets. 

In such systems, the presence of the known transiting planets can provide some constraint on the location of the non-transiting perturber. For example, KOI-1781.01/Kepler-411c has a strong sinusoidal TTV corresponding to a super-period of 1290 days. With a period of 7.88 days, Kepler-411c is too far from the other planets with periods of 3.01 and 58.02 days for them to be the cause of these TTVs; the closer planet would be related by a period ratio of 2.62 and is not consistent with the observed super-period. A near-resonant non-transiting perturber planet is the most likely cause. However, a planet at the interior 2:1 resonance would have a 3.9-day period, potentially too close to the 3-day planet. Even when the TTVs cannot be ascribed to known planets, multi-transiting systems still provide insight unavailable in singly-transiting systems. 

\section{Conclusions} \label{conclusion}

Kepler TTVs have emerged as an important way for measuring planetary masses and densities and identifying non-transiting planetary perturbers. H+16 provided statistical metrics that could be used to identify a list of TTVs for further study. Though TTVs usually take the form of the sum of periodic signals, the variety of possible outcomes suggested that these statistical metrics should be validated by visual inspection. Furthermore, significant additional TTV data from \citet{2015arXiv150400707R} prompted additional investigation. We make visualizations of these TTVs available to the community at \url{http://haumea.byu.edu/kanettv/\#.pdf} where \# is replaced by the KOI number. 

With this motivation, we present homogeneous visual ratings of 5930 KOIs based on signal cleanliness, the quality of a quadratic+sinusoid fit, the strength of periodic signals, and overall interestingness (Table \ref{table:meta_data_table}). Though these are subjective in nature and did not exhibit much dynamic range (Table \ref{table:distribution_table}), they are directly correlated in expected ways with the statistical metrics of H+16 (see Figure \ref{fig:int_poly_scat}). Many KOIs that were not studied in detail by H+16, but which have significant signals, have been identified by our ratings. We focus herein on the results for planets, but we have ratings for 2144 false positives that could be of value for understanding Eclipsing Binaries (see Table \ref{table:eb_table}) and other false positives and false alarms. 

Our ratings emphasize that there are hundreds of planets with weak-but-real TTVs that remain to be studied in detail. In particular, we note here two methods which raise the value of weak (and strong) TTVs: the use of Short Cadence data and the use of photodynamical modeling. Kepler Short Cadence data, where photometric observations were returned every minute instead of every 30 minutes, improve the precision of Transit Times by resolving the ingress/egress time, especially for small planets. Several empirical, analytical, and numerical estimates show that Short Cadence data leads to improved TTV measurements, especially for small planets, regardless of the SNR per transit \citep[e.g.][]{2012Sci...337..556C}. The Kepler TTV/Multis Working Group ensured that hundreds of the strongest TTV candidates were observed in Short Cadence, implying great potential for improved analyses of what we classified as "weak" TTVs. 

Another area of improvement is the use of photodynamical models. These transcend TTV analyses and allow for self-consistent investigation of planets with SNR-per-transit too small for rigorous TTVs. Photodynamical modeling extracts all the information possible without resorting to summary metadata like transit times and is essential for systems where duration and/or depth variations may be important (Section \ref{tdvtpv}; \citet{RH10}). An example is the Kepler-444 system, which we rate as having some strong TTVs, but where a photodynamical model was able to put scientifically-valuable compositional constraints on two of the planets \citep{2017ApJ...838L..11M}. Our identification of several hundred weak TTVs indicates that the photodynamical approach could be profitable on much larger number of systems than have currently been analyzed. 

In addition to the use of Short Cadence and/or photodynamical models, our work shows the viability of combining hundreds of TTV constraints in an ensemble analysis of the mass-radius relation and/or planetary architectures. 

The presence of strongly detectable TTVs that lead to a high Overall rating depends on a variety of physical, orbital, and observational factors. Even so, we investigate the demographics of planets (CONFIRMED/CANDIDATE from the Exoplanet Archive) with strong TTVs using the entire sample of planets as a control group (see Figure \ref{fig:perradttv}. For the most part, we identify trends that are consistent with observational bias. One clear exception is a clear dearth of TTV signals around planets with periods $\lesssim$3 days, including both Hot Jupiters and Hot Earths. These results are consistent with many other studies that show that this short-period population is unique. There is also evidence that the intrinsic frequency of TTVs among large planets is lower than among smaller planets. 

Strong TTVs are about 1.6 $\pm$ 0.2 times as frequent (on a planet-by-planet basis) in systems with three or more transiting planets than in systems with only one or two known transiting planets. This can be explained with a hypothesis that systems which are more tightly-packed are both more likely to have more planets transit and more likely to have detectable TTVs. More work will be required to determine whether this is due to a two-population model as suggested by the so-called Kepler dichotomy \citep[e.g.,][]{2012ApJ...758...39J, 2016ApJ...832...34M, 2016ApJ...822...54D}. A key component of any model that attempts to describe the frequency of TTVs will be an accurate representation of the near-resonant period ratios, where TTVs are strongly enhanced (Figure \ref{fig:perratio}). However, there is not a one-to-one correspondence between near-resonant period ratios and strong TTVs. 

We identify, where possible, which planets are responsible for causing strong TTVs. About 1/4 of planets in multi-transiting systems have TTVs that cannot be attributed to the known planets, but these can be readily ascribed to non-transiting planets which are expected, even in multi-transiting systems. 

In two appendices, we explore the TTVs from Kepler Data Release 25 and provide ratings for the strongest Transit Duration and/or dePth Variations (TDVs/TPVs) from H+16. We identify some of the most interesting results. 

Overall, we hope that our analysis provides a launching off point for future investigations. We provide plots and ratings to identify individually interesting systems that deserve deeper analysis and indicate areas where future modeling of planetary architectures could employ TTVs as a constraint. Deep analyses of TTVs hold significant promise in better understanding the formation, evolution, architectures, frequency, and habitability of planetary systems.

\begin{acknowledgments}

We thank the \ik Team for their outstanding work. We thank many for their comments, suggestions, and contributions including Kat Deck, Eric Ford, Sam Hadden, Daniel Jontof-Hutter, and $Kepler$'s TTV/Multis Working Group. We acknowledge the support from the Florida Institute of Technology Fall 2015 SPS1020 class. MK, DR, and XF acknowledge support from NASA Origins of Solar Systems grant \# NNX14AI76G. This research has made use of the NASA Exoplanet Archive, which is operated by the California Institute of Technology, under contract with the National Aeronautics and Space Administration under the Exoplanet Exploration Program. The citations in this paper have made use of NASA's Astrophysics Data System Bibliographic Services.

\end{acknowledgments}

\appendix

\section{DR25 Planet Candidates with Interesting TTVs}
\label{DR25}
After the completion of the visual analysis described in the main text, the \ik mission completed its final planet search and characterization, known as ``Data Release 25'' or DR25 \citep{2017arXiv171006758T}. Our analysis ends with KOI-5978, but the final catalog goes to KOI-8297, though most new KOIs are false positives or false alarms. 

\citet{2015arXiv150400707R} calculated TTVs for all KOIs as part of the Kepler pipeline. Similar work has been performed on the DR25 available at \url{https://exoplanetarchive.ipac.caltech.edu/docs/Kepler\_KOI\_docs.html}, but transit times are only calculated for those KOIs where accounting for TTVs is important for fitting the transit light curve. See KSCI-19113-001 for more details. 

DR25 transit times were provided for 296 KOIs. Most of these have existing TTV data from RT15 or H+16. Using the new DR25 transit times, we made TTV plots like Figure \ref{fig:TTV_full}. Like the other TTV plots, these are available at \texttt{http://haumea.byu.edu/kanettv/\#DR25.pdf} where \# is the KOI number, e.g., \url{http://haumea.byu.edu/kanettv/351.07DR25.pdf}. 

Entirely new TTV calculations are available for 16 KOIs, of which 4 are planet candidates (using the DR25 disposition). Most of these are new KOIs in existing systems. 

Some new signals of interest from planetary candidates: 
\begin{itemize}
\item KOI-351.07 in the 7-planet Kepler-90 system shows significant TTVs.
\item KOI-520.04/Kepler-176e shows evidence of chopping which could be due to KOI-520.03/Kepler-176d. 
\item KOI-1573.02 has a significant periodic signal that is probably due to a non-transiting planet. 
\end{itemize}
Many false positives show interesting signals, but we do not enumerate them here. 

It is important to remember that the Kepler pipeline dispositions are not always able to correctly handle planets with TTVs. This is because the ``robovetter'' \citep{2017ksci.rept....1C} uses the folded transit shape as part of its assessment and planets with TTVs that are comparable to the duration of the transit will have folded transit shapes that are inconsistent with a planetary transit \citep[e.g.,][]{2011MNRAS.417L..16G}. Hence, the disposition of "FALSE POSITIVE" due to "Not Transit-Like" shape is sometimes given to planets with TTVs (e.g., KOI-8151). This explains why several confirmed planets have ``false positive'' dispositions. It is beyond the scope of this paper to systematically identify these cases, but our results can be helpful in finding TTVs where this might be a concern.

\section{Investigation of Duration and Depth Variations} \label{tdvtpv}

As part of their procedure to fit transit light curves for TTVs, H+16 were also able to measure durations and depths for transits where the SNR per transit was greater than 10 and the duration longer that 1.5 hours. For 779 KOIs, they measured 69,914 times, durations, and depths. 
The durations and depths were then analyzed for any Transit Duration Variations (TDVs) or Transit dePth Variations (TPVs) by searching for significant periodicities and by identifying possible long-term trends (slopes). Unlike TTVs, where slopes are subtracted out to find the best-fit orbital period, long-term changes in durations and depths are meaningful. The possibility and dynamical implications of observing these changes has been proposed by many authors \citep[e.g.,][]{2002ApJ...564.1019M,2009ApJ...698.1778R,2013ApJ...777....3N}. Astrophysical variations include precessing orbits (which changes on-the-sky velocities and/or impact parameters) and/or stellar variability (which changes the depth of the transit).

H+16 did not present an analysis of their statistical results for TDVs/TPVs, so we provide a brief discussion here. 

At \url{ftp://wise-ftp.tau.ac.il/pub/tauttv/TTV/ver\_112}, they host \texttt{TDV\_statistics.txt} and \texttt{TPV\_statistics.txt}. Using these files, we identify as potentially interesting variations any KOI with a periodogram p-value of less than 0.0006 or with an estimated slope that exceeds 3.5 times the estimated error in the slope. Using these thresholds, 28 (79) KOIs were identified as having potentially interesting TDVs (TPVs) for a total of 97 KOIs, including 10 in both categories. Of these, 76 belong to planets or planet candidates. 

Using the data from Table 2 in H+16, we made simple plots of the TTVs, TDVs, and TPVs for these 97 systems. Only data that did not have any overlap or outlier flags were included (see H+16), but there were still many outliers. The plots showed the core 90\% of the data with error bars in order to clearly identify the signals. We then performed a simple visual inspection to help identify the most interesting cases. The file showing the TDVs and TPVs that was used for inspection is at \url{http://haumea.byu.edu/kanettv/CheckTDVTPV.pdf}. 

The inspection placed every potentially interesting signal into one of 3 categories: noise/spurious, a slope worth further investigation, and an oscillation worth further investigation. In particular, one or more of the following values were assigned for all 97 KOIs. 

\begin{itemize}
\item 1 $\rightarrow$ TDV noise/spurious 
\item 2 $\rightarrow$ TDV slope worth further investigation
\item 4 $\rightarrow$ TDV oscillation worth further investigation
\item 10 $\rightarrow$ TPV noise/spurious
\item 20 $\rightarrow$ TPV slope worth futher investigation
\item 40 $\rightarrow$ TPV oscillation worth further investigation
\end{itemize}

These values were summed to provide the final \texttt{TDPV} value in Table \ref{table:meta_data_table}. KOIs which had measured TDVs and TPVs, but which did not reach the level of potentially interesting were assigned a flag value of 0. KOIs with no TDV/TPV measurements in H+16 were assigned a \texttt{TDPV} value of -1. 

As with the TTV ratings, these results are somewhat subjective and are only provided as a guide for additional investigation. However, we found that 26 (of 28) KOIs showed interesting TDV signals and 20 (of 79) KOIs showed interesting TPV signals (with 3 overlaps). 

The vast majority of strong signals where we did not consider the signal worth additional investigation were those with depth oscillations with the same period as \ik itself. The observation pattern of \ik can introduce a dilution or contamination that is different every quarter, but repeats every $\sim$372 days as the spacecraft returns to its original orientation. This is a well-known effect and most periodic TPVs were afflicted by this instrumental effect and therefore received scores of 10 (noise/spurious). 

We comment briefly on some of the interesting systems. Some of the systems with significant duration and/or depth variations were already known, such as KOI-13.01/Kepler-13b \citep{2012MNRAS.421L.122S},  KOI-142.01/Kepler-88b \citep{2013ApJ...777....3N}, KOI-119.02 \citep{2017AJ....153...45M}, KOI-1546.01 \citet{2015ApJ...807..170H}, KOI-3853.01 \citep{2015A&A...576A..88L}, and KOI-824.01/Kepler-693b \citep{2017AJ....154...64M}. Most of these have been studied with photodynamical models, revealing the physical nature (oblateness, high mutual inclination, etc.) of the observed variations. 

Still, there are many new cases worth additional investigation, including the following examples. 
\begin{itemize}
\item KOI-1.01/Kepler-1b/TrES-2b has secular depth changes (though this target is saturated, which makes accurate photometry difficult). 
\item KOI-377.01/Kepler-9b and KOI-377.02/Kepler-9c both show clear TDV slopes in opposite directions which may provide additional insight into this system. Systems with strong TTVs like Kepler-9 sometimes demonstrate oscillating TDVs with the same period and similar phase, but the TDVs seen in Kepler-9 are long-term slopes, indicating a true change in inclination due to nodal precession, but further modeling is recommended. KOI-137.01/Kepler-18c and KOI-137.02/Kepler-18d may show a similar trend. 
\item KOI-1426.02/Kepler-297c and KOI-1426.03 show coherent TPV slopes, in addition to very strong TTVs. 
\item KOI-3678.01 is a 160-day Jupiter that shows TTVs, TDVs, and TPVs which all appear significant. 
\end{itemize}

The TDV/TPV signals identified as potentially interesting by \texttt{TDPV} represent strong signals likely to be real. There are certainly weaker TDV/TPV signals that were not identified which could also provide useful constraints on physical and orbital properties of the planets using photodynamical modeling.

\bibliographystyle{aasjournal}
\bibliography{all}

\end{document}